\documentclass[a4paper, UKenglish]{oasics-v2021}

\pdfoutput=1 
\hideOASIcs 

\title{Accelerator-driven Data Arrangement to Minimize Transformers Run-time on Multi-core Architectures}

\titlerunning{Data Arrangement for Transformers} 

\author{Alireza Amirshahi}{École Polytechnique Fédérale de Lausanne (EPFL), Switzerland}{alireza.amirshahi@epfl.ch}{https://orcid.org/0000-0002-6229-9103}{}

\author{Giovanni Ansaloni}{École Polytechnique Fédérale de Lausanne (EPFL), Switzerland}{giovanni.ansaloni@epfl.ch}{https://orcid.org/0000-0002-8940-3775}{}

\author{David Atienza}{École Polytechnique Fédérale de Lausanne (EPFL), Switzerland}{david.atienza@epfl.ch}{https://orcid.org/0000-0001-9536-4947}{}

\authorrunning{A. Amirshahi, G. Ansaloni, and D. Atienza} 



\category{} 

\relatedversion{} 

\supplement{Source Code: \href{https://github.com/gem5-X/TiC-SAT}{https://github.com/gem5-X/TiC-SAT}}

\funding{This research was partially supported by EC H2020 FVLLMONTI project (GA No. 101016776) and by the ACCESS – AI Chip Center for Emerging Smart Systems, sponsored by InnoHK funding, Hong Kong SAR.}

\nolinenumbers 

\EventEditors{}
\EventNoEds{0}
\EventLongTitle{}
\EventShortTitle{}
\EventAcronym{}
\EventYear{}
\EventDate{}
\EventLocation{}
\EventLogo{}
\SeriesVolume{}
\ArticleNo{}

\begin{document}

\maketitle

\begin{abstract}
The increasing complexity of transformer models in artificial intelligence expands their computational costs, memory usage, and energy consumption. Hardware acceleration tackles the ensuing challenges by designing processors and accelerators tailored for transformer models, supporting their computation hotspots with high efficiency. However, memory bandwidth can hinder improvements in hardware accelerators. Against this backdrop, in this paper we propose a novel memory arrangement strategy, governed by the hardware accelerator's kernel size, which effectively minimizes off-chip data access. This arrangement is particularly beneficial for end-to-end transformer model inference, where most of the computation is based on general matrix multiplication (GEMM) operations. Additionally, we address the overhead of non-GEMM operations in transformer models within the scope of this memory data arrangement. Our study explores the implementation and effectiveness of the proposed accelerator-driven data arrangement approach in both single- and multi-core systems. Our evaluation demonstrates that our approach can achieve up to a 2.8x speed increase when executing inferences employing state-of-the-art transformers.
\end{abstract}

\section{Introduction}
In recent years, the rise of transformer models has significantly impacted the field of artificial intelligence, being used in various domains such as natural language processing, machine translation, speech recognition, and computer vision~\cite{devlin2018bert, dosovitskiy2020image, hsu2021hubert}. As a result, the demand for ever more powerful transformers has emerged, leading to the development of increasingly large and complex architectures.

The substantial size and complexity of transformers have raised new challenges in the ability to cope with their computational costs, memory usage, and energy requirements.  In fact, the increasing complexity of these models must be carefully managed to contain the ensuing processing needs, which can lead to unacceptable operating costs \cite{zhong2023transformer}. 

Several works have addressed these challenges \cite{kim2023full}. The first notable avenue is that of hardware acceleration, that is, the design of processors and accelerators specifically tailored for transformer models~\cite{qi2021accelerating, you2023vitcod, li2020ftrans}. Such specialized hardware aims to significantly reduce computation time by distributing the calculations across multiple smaller processing units within the accelerators.

The second strategy to optimize the execution of the transformer model is the enhancement of the locality of the data. This approach involves maximizing the reuse of data already fetched from slow and energy-hungry off-chip memory. Given the bandwidth limitations of accessing off-chip memory, reducing reliance on this memory and improving data locality can significantly enhance performance in terms of energy and time efficiency. Methods to improve data locality include employing cache hierarchies as on-chip memory, using loop tiling to divide computational workloads into smaller segments~\cite{lam1991cache}, and adopting in-memory or near-memory computation techniques~\cite{khoram2017challenges, eggermann2023sixteen}. 

Although reusing data fetched from memory can considerably improve transformer inference efficiency~\cite{amirshahi2023tic}, the sequence in which the data are retrieved remains crucial, especially when using a hardware accelerator. This issue forms the basis of our research focus. In this context, all multi-dimensional data structures, e.g., matrices, must be converted into one-dimensional arrays for storage, a process we term \emph{data arrangement}. The arrangement of data is essential in systems with hardware accelerators and memory hierarchies. In this paper, we propose for such systems a memory-data arrangement that aligns with the processing unit size of hardware accelerators. In this data arrangement, when a data block is loaded into the accelerator, a contiguous block can simultaneously be pre-fetched from off-chip memory because the data arrangement is synchronized with the processing sequence. This approach leads to faster data retrieval for subsequent accelerator loads by handling as many contiguous memory accesses as possible. 

Key contributions of this paper include the following:

    \begin{itemize}
        \item The introduction of a memory data arrangement that coordinates with the size of the hardware accelerator, facilitating synchronized data access and storage.
        \item A comprehensive analysis of implementing this memory data arrangement in an end-to-end transformer model, highlighting its advantages in general matrix multiplication (GEMM) operations with minimal added overhead for non-GEMM processes.
        \item The demonstration of the feasibility of deploying this memory data arrangement in both single- and multi-core hardware accelerators.
        \item Extensive full-system evaluations, revealing that our proposed memory data arrangement can greatly speed-up inference in transformer models, e.g. by up to 2.8x in a single-core system.
    \end{itemize}

The remainder of this paper is structured as follows. Section~\ref{sec:background} explains the background and related work, providing key information on transformers, hardware accelerators, and existing studies on memory data arrangements. Section~\ref{sec:methods} describes our methodology, introducing the accelerator-driven data arrangement and its impact on transformer models. In Section~\ref{sec:experiments}, we present our experimental setup and results, showcasing the effectiveness of the proposed memory data arrangement in enhancing the performance of the transformer model. Finally, Section~\ref{sec:conclusion} concludes the paper, summarizing our findings and contributions.

\section{Background and Related Works}
\label{sec:background}

\subsection{Transformers}
Transformer models consist of multiple interconnected components that generate outputs based on weighted inputs according to a self-attention technique.  The self-attention mechanism enables the model to dynamically adjust the weights of different inputs, thus allowing it to capture dependencies between features (e.g., words in case of machine translation), even when distant within the input data. The original version of the transformer model~\cite{vaswani2017attention} presents an encoder-decoder structure that has proven effective for tasks that require sequence-to-sequence transformations. Several derivatives have been developed, primarily utilizing the encoder portion of the model. The encoder itself has the ability to learn robust high-level representations from input data, which makes these models particularly effective in a wide range of tasks, such as sentiment analysis, question answering, image recognition, and more~\cite{devlin2018bert, dosovitskiy2020image}.  In encoder-focused models, the encoder part maintains a structure similar to~\cite{vaswani2017attention}, while the decoder part is often simplified to a single- or two-layer linear component. Consequently, in these applications, optimizations focus on the encoder layers, which comprise the majority of the computational workload. We follow the same approach, aiming to minimize the run-time for encoder-focus transformer models.

\begin{figure}[t]
    \centering
    \includegraphics[width=0.5\linewidth]{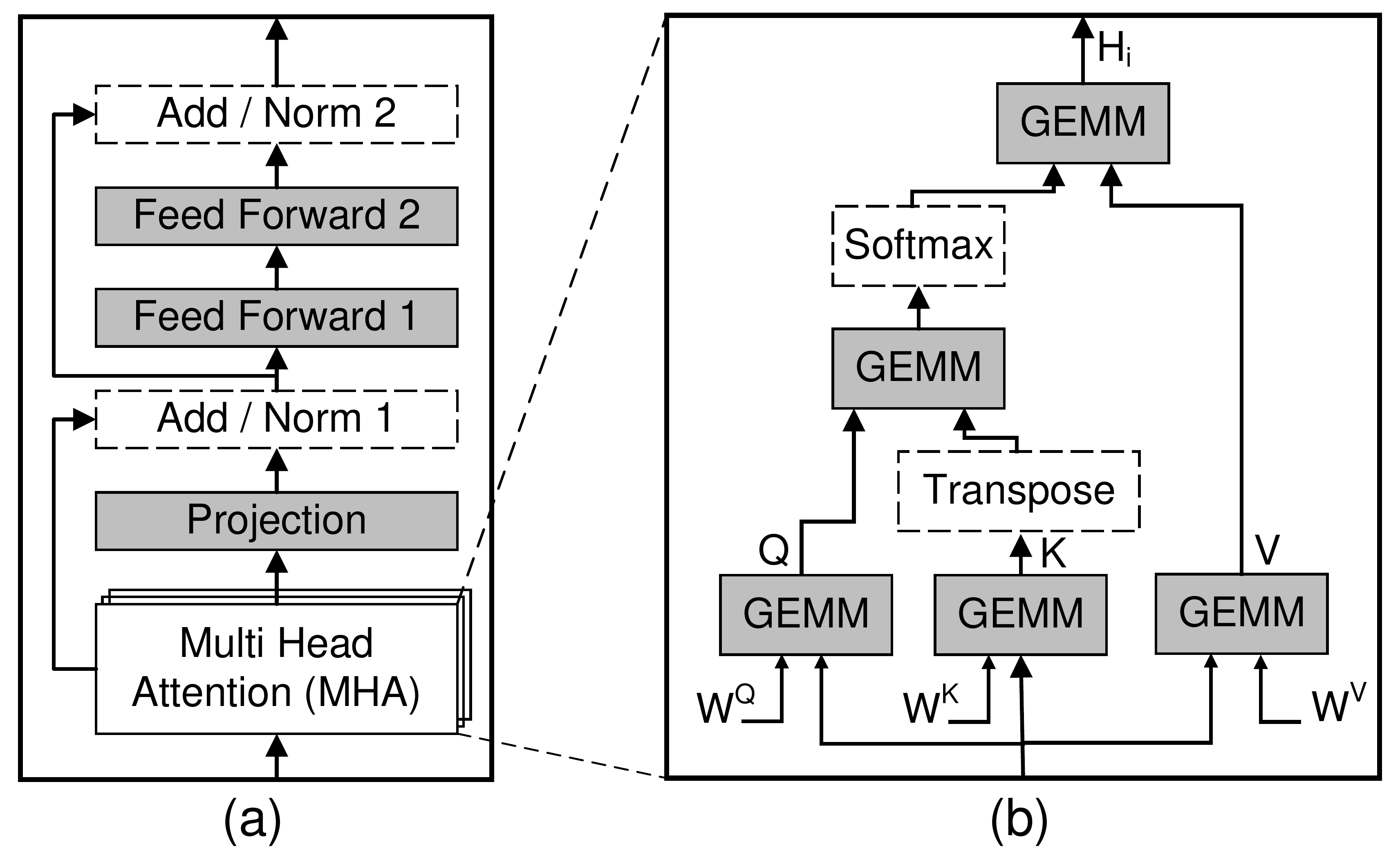}
    \caption{(a) Overall structure of a transformer encoder layer, and (b) detailed representation of a single attention head with its components. The GEMM-based components within the encoder layer and attention head are shaded in grey. Components corresponding to non-GEMM operations have dashed frames.}
    \label{fig:transformer}
\end{figure}

 The architecture of an encoder layer within a transformer model is depicted in Figure~\ref{fig:transformer}a. The Multi-head Attention component takes as input a matrix $X$ with a number of rows equivalent to the sequence length and a number of columns corresponding to the model dimension. This input matrix is applied to $h$ distinct heads, each having a unique set of weight matrices, which generates unique output values from the same input. In more detail, the input matrix $X$ is multiplied by $W^Q_i$, $W^K_i$, and $W^V_i$, where $i \in {0, .., h-1}$ represents the head index. The resultant matrices are $Q_i$, $K_i$, and $V_i$, which correspond to Query, Key, and Value matrices, respectively. These multiplications are illustrated in Figure~\ref{fig:transformer}b as \emph{GEMM} operations.

Subsequently, the next component involves transposing $K_i$ and multiplying it by $Q_i$. Following this, a non-linear Softmax operation is applied to the results, and the output is multiplied by $V^i$. 
Overall, the output of this single-head attention component is computed as follows:

\begin{equation}
H_i = \text{Softmax}( \frac{Q_i \times K_i^T} {\sqrt{d_q}}) \times V_i \;\; i \in {0, 1, \dots, h-1}
\label{eq:single_head}
\end{equation}

In the above equation, $d_q$ denotes the dimension of the Query, Key, and Value matrices. The softmax operation subsequently scales the matrix values within the range $[0,1]$. $H_i$ represents the output of the $i$-th head and, along with the outputs of all other attention heads, serves as input for the projection component.

As illustrated in Figure~\ref{fig:transformer}a, the outputs from the heads are concatenated and then subjected to a projection. The projection result is then routed to the Add/Norm component, which performs a non-linear operation. Following this, the output is passed through two feed-forward components. The final stage in the process involves a second Add/Norm operation.

In summary, the core of a transformer encoder layer is predominantly composed of General Matrix Multiply (GEMM) operations. Thus, enhancing the efficiency of GEMM operations is key to accelerating transformer run-time. The proposed method for GEMM acceleration using the data arrangement and the overhead associated with non-linear operations are discussed in detail in Section~\ref{sec:methods}.
\subsection{Hardware Accelerators}
\label{sec:hardware_accelerator}

\subsubsection{Accelerators for GEMM Operations}
This paper focuses on two predominant categories of accelerators used in GEMM operations: Systolic Arrays (SA) and Single Instruction Multiple Data (SIMD) architectures. Both types are characterized by their integration of data storage registers and processing elements for computation.

\begin{figure}[t]
  \centering
  \begin{subfigure}{0.49\textwidth}
    \centering
    \includegraphics[width=\linewidth]{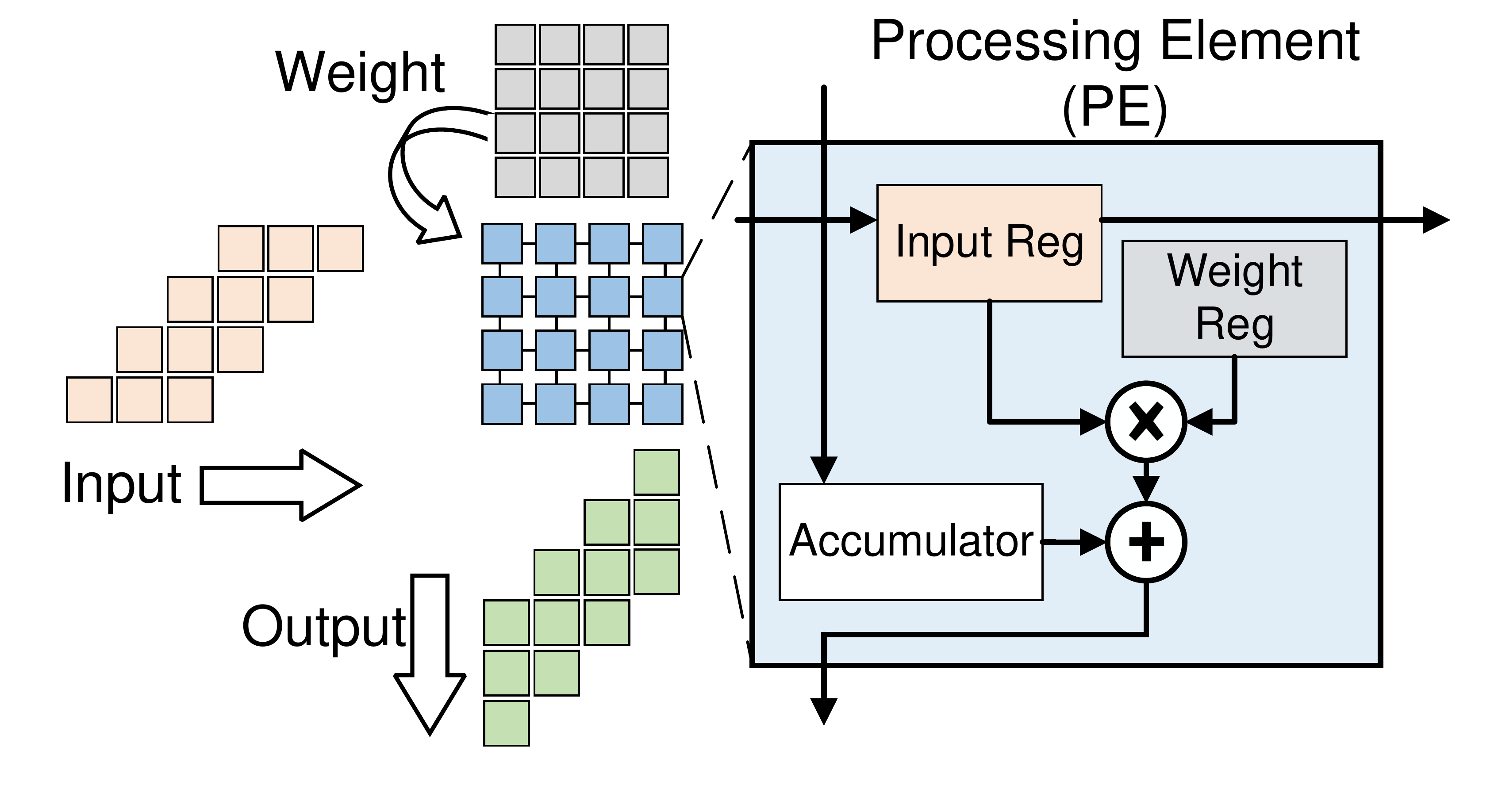}
    \caption{}
    \label{fig:systolic_array}
  \end{subfigure}
  \begin{subfigure}{0.49\textwidth}
    \centering
    \includegraphics[width=\linewidth]{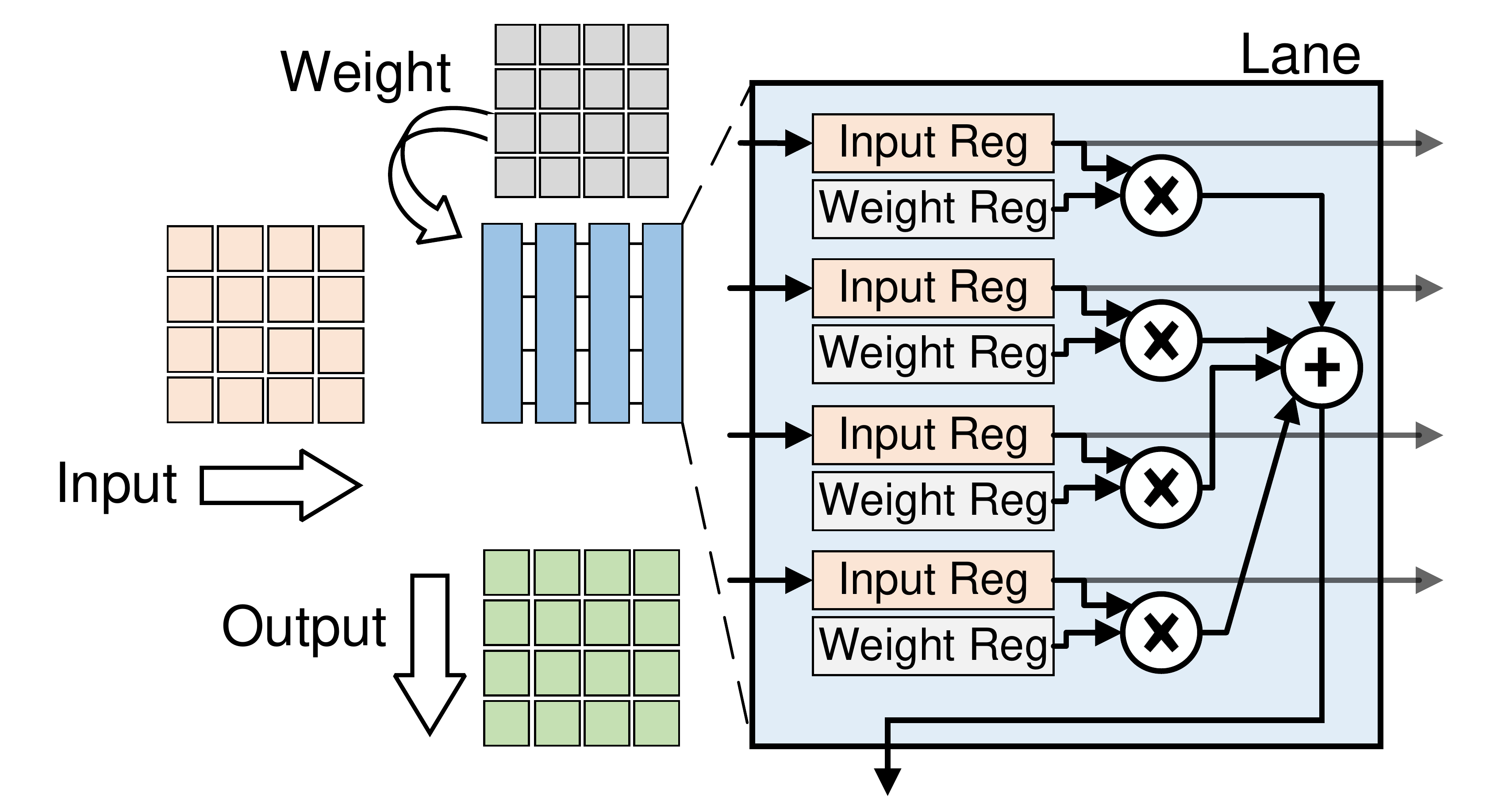}
    \caption{}
    \label{fig:simd_arch}
  \end{subfigure}
  \caption{GEMM acceleration using (a) Systolic Arrays and (b) SIMD architectures.}
  \label{fig:accelerators}
\end{figure}

\textbf{Systolic Arrays} consist of a network of Processing Elements (PEs) arranged to process input streams into output streams. Each PE is equipped with arithmetic and storage units, namely an adder, a multiplier, and three registers. Figure~\ref{fig:systolic_array} shows a 4x4  SA architecture. In this configuration, input and output data propagate through the array in orthogonal directions (for instance, inputs moving left-to-right and outputs top-to-bottom). Weights are preloaded into the PEs before computation begins. During operation, the inputs shift unaltered from one PE to the next with each clock cycle, while the outputs accumulate results from the multiplications of the inputs and weights. 

\textbf{SIMD} architectures, as shown in Figure~\ref{fig:simd_arch}, comprise multiple computing lanes that simultaneously execute the same operation on different data. Each lane performs a dot-product operation. To execute dot-product operations in parallel, the SIMD system loads weights into each lane's registers and processes the input matrix through these lanes to compute the output.

In both SA and SIMD architectures, the configuration of processing elements or lane arrays can be referred to as the accelerator kernel. The term \emph{kernel size} denotes the number of PEs in each row or lane. Our research aims to optimize the memory data arrangement aligning with this kernel size to enhance the efficiency of GEMM operations in these accelerators for a transformer application.

\subsubsection{Tiling for GEMM Operations}
The size of the matrices involved in transformer models greatly exceeds the accelerator kernel size. Hence, to use the accelerators when computing large GEMMs, matrices must be tiled; that is, partitioned into smaller submatrices. Partial GEMMs are then performed on tiles, increasing locality and data reuse in accelerator kernels. Finally, the results are aggregated with element-wise additions.

\begin{figure}[t]
    \centering
    \includegraphics[width=\linewidth]{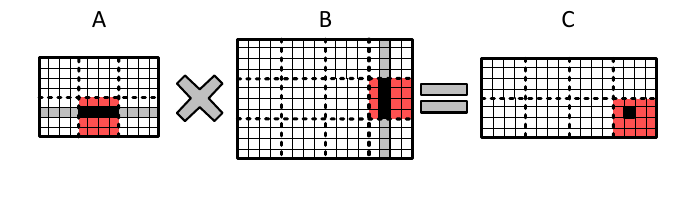}
    \caption{Tiled matrix multiplication $A \times B = C$, considering tiles of size 4x4 elements.}
    \label{fig:tiling}
\end{figure}

Figure~\ref{fig:tiling} illustrates this concept with matrices A and B undergoing multiplication to produce matrix C. Within each of the three matrices, a 4x4 tile is highlighted. Multiplying the highlighted tiles into matrices A and B gives a partial result in the corresponding area of matrix C. For example, the black-marked value in matrix C is partially derived by multiplying the black row in A by the corresponding black column in B. Note that the tiling yields only partial values; however, by sliding the tiles and accumulating these partial results through element-wise addition, the final values in the output matrix C can be derived.

\subsection{Related Works}

\subsubsection{Memory Data Arrangements}
Rearranging the layout of the matrix elements can greatly speed up GEMMs, as shown in~\cite{herrero2006using}. This research focused on the execution of a singular General Matrix Multiply (GEMM) operation, where the matrices were arranged in a square block format where the matrices are stored as square submatrices column by column. 
Building on this concept, the authors in \cite{ferry2022increasing} applied the block matrix layout to operations beyond GEMM, such as single convolution and Gaussian blur. Their research demonstrated that with the implementation of an optimized memory layout, they were able to fully utilize the memory capacity. 

Our proposed method diverges from the approaches in \cite{herrero2006using} and \cite{ferry2022increasing} by employing block-wise memory arrangement in the context of hardware accelerators. We strategically align this kernel size with the blocks used for data storage in memory. Furthermore, we extend the application of this memory arrangement to a complete transformer application, which comprises heterogeneous layers, as opposed to a single operation among matrices. This approach allows us to explore the potential benefits and efficiencies of aligning the memory arrangement with the hardware accelerator size in realistic computational scenarios.

\subsubsection{Hardware Accelerators for Transformer Models}
A comprehensive survey of hardware accelerator implementations for transformer models is provided by~\cite{kim2023full}, discussing state-of-the-art hardware accelerator frameworks and potential research directions in transformer hardware acceleration.
One notable accelerator design, proposed by~\cite{peng2021accelerating}, introduces an accelerator on FPGAs, which achieves lower latency and higher energy efficiency compared to CPU, GPU, and prior FPGA-based accelerators. \cite{yang2022efa} presents a hardware accelerator architecture that utilizes a configurable matrix computing array and on-chip buffers on the FPGA. However, these two papers study accelerators in isolation, as opposed to considering them as part of an overall computing system. On the contrary, here we adopt a system-level view, investigating the interplay between memory hierarchy, accelerator design, and system performance and proposing strategies to minimize data access at lower memory levels.
 
Our approach is related to \cite{amirshahi2023tic}, which also addresses the challenge of minimizing data transfers by introducing tightly coupled small-scale systolic arrays. These are integrated as custom functional units and are governed by dedicated ISA extensions. The ensuing data reuse results in significant application-wide speed-ups. However, the work in \cite{amirshahi2023tic} does not consider optimization of the memory data arrangement as a hardware-friendly optimization strategy.

\section{Methodology}
\label{sec:methods}
\subsection{Aligning memory arrangement with the accelerator kernel size}
\label{sec:block-wise-mem}

Our proposed technique, called Block-Wise Memory Arrangement (BWMA), further enhances the benefits of tiling by optimizing the data arrangement in the off-chip memory. Before illustrating BWMA, we contrast it with a conventional approach for storing two-dimensional matrices in (linear) memory, which is named herein Row-Wise Memory Arrangement (RWMA).

\subsubsection{ Row-Wise Memory Arrangement (RWMA)}
Figure~\ref{fig:abw_method}a illustrates an 8x8 matrix containing various patterns as rows and different colors in the columns. To store this matrix in memory, it must be converted to a one-dimensional array. As demonstrated in Figure~\ref{fig:abw_method}c, RWMA stores the matrix row by row in memory, treating them as 1-D arrays. When this matrix is incorporated into a GEMM operation via the tiling technique, irregular non-sequential sections of the 1-D memory are accessed to retrieve the tile. Following the GEMM operation, storing the resultant tile back into memory also necessitates access to non-sequential memory regions. 

\subsubsection{Block-Wise Memory Arrangement (BWMA)}
\begin{figure}[t]
    \centering
    \includegraphics[width=0.75\linewidth]{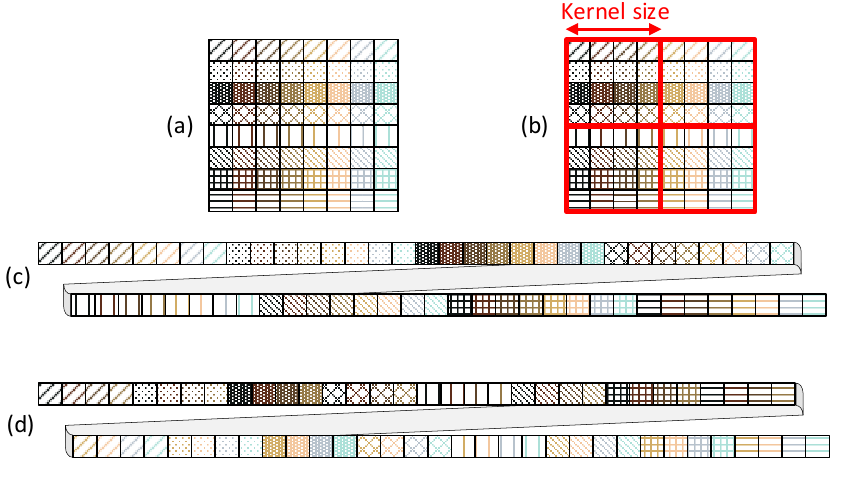}
    \caption{Comparison of conventional row-based memory arrangement (a, c) and the proposed block-wise memory arrangement (b, d) for an 8x8 matrix.}
    \label{fig:abw_method}
\end{figure}

 As shown in Figure~\ref{fig:abw_method}b, our proposed approach partitions weights, input, and output matrices into smaller blocks. The size of the blocks in this partitioning is as large as the size of the kernel in the hardware accelerator. As discussed in Section~\ref{sec:hardware_accelerator} and shown in Figure~\ref{fig:accelerators}, the size of the kernel is defined as the number of PEs in each row (in SAs) or lane (in SIMDs). The blocks, as opposed to rows, are then stored in sequence as 1-D arrays in memory (see Figure~\ref{fig:abw_method}d). 

 BWMA ensures that GEMM operations fetch data from sequential sections of the 1-D array, as stored in memory. This arrangement is particularly advantageous for systems having a multi-level memory hierarchy, i.e., having a large/slow main memory and multiple layers of progressively faster/smaller caches. In these systems, BWMA  allows the expected contiguous data to be pre-fetched correctly into caches and, ultimately, in the hardware accelerator. In fact, BWMA allows for a higher degree of data reuse across the entire memory hierarchy, reducing costly accesses to lower memory levels.
 
\subsection{Impact of BWMA in transformer models}
\label{sec:bw_mem:non-gemm}

In practical applications, even if the input matrix for a transformer model is structured using RWMA, the transition to-from RWMA and BWMA, is only necessary at the start and at the end of the entire transformer computation process for the input and output matrices. Importantly, there is no necessity to rearrange all the intermediate matrices produced by the GEMM operations, as these can be supplied to the subsequent layer in a block-wise fashion. 

Given the multi-layered structure of the transformer and the composition of several components within each transformer layer, as discussed in Section~\ref{sec:background}, the overhead associated with the RWMA $\leftrightarrow$ BWMA transitions of input and matrices is insignificant in comparison to the overall transformer computation. We observed that they consume only 0.1\% of the total execution time for an entire 12-layer transformer model.

\begin{figure}[t]
    \centering
    \includegraphics[width=0.7\linewidth]{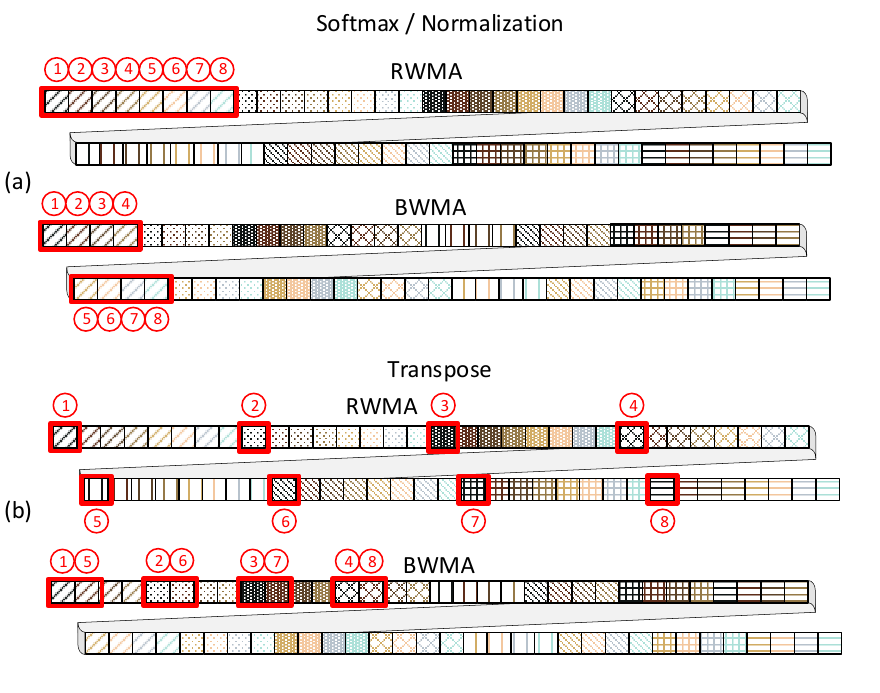}
  \caption{Non-GEMM operation access pattern in RWMA and BWMA for (a) Softmax/Normalization and (b) Transpose layers. The first accessed eight elements are represented in each experiment.}
  \label{fig:non_gemm_overhead}
\end{figure}

Through the transformer model,  non-GEMM functions are also influenced by BWMA. The transpose and activation functions are similar, from a computational perspective, in conventional RWMA and block-based BWMA. Softmax and Normalization require proper indexing of blocks in the BWMA case, which has an impact on their run-time. However, this effect is negligible with respect to the run-time of GEMM operation, as overall non-GEMM operations take at most 13.5\% of the execution time (even when GEMM is hardware accelerated), as shown in Section~\ref{sec:results:run-time}. The impact of BWMA on non-GEMM functions are explained as follows:

\textbf{Softmax:} The softmax layer computes the exponential function for each element in a matrix row and normalizes the result, producing probabilities. The access pattern to elements in a row differs between RWMA and BWMA. Figure~\ref{fig:non_gemm_overhead}a illustrates the first eight elements accessed during reading in RWMA and BWMA. As shown, the data is read consequently in RWMA, whereas in BWMA the reading is a non-sequential pattern, introducing overhead in this data arrangement. The processed data is written back to the same matrix position, leading to additional overhead in BWMA compared to RWMA for the write-back operation.

\textbf{Normalization:} This process involves computing a row's element sum, and then calculating the average. Subsequently, the variance is determined using the calculated average and its square root yields the standard deviation. Each row element is normalized by subtracting the average and dividing by the standard deviation. The memory access pattern of the normalization function is identical to the softmax function, as both operate on a row-by-row basis. Therefore, as demonstrated in Figure~\ref{fig:non_gemm_overhead}a, this function presents a slight overhead for BWMA compared to RWMA.

\textbf{Transpose:} This operation requires swapping elements in the matrix to change its orientation. The access pattern in this operation differs notably between RWMA and BWMA. Figure~\ref{fig:non_gemm_overhead}b shows the initial eight elements accessed in this function for both arrangements. During data reads, neither method accesses the data sequentially; however, BWMA exhibits better data locality. Note that this figure only represents data accessed during reading.  Writing back the data in the memory after transpose is sequential for both data arrangements.

\textbf{Activation:} This function is an element-wise operation. Therefore, the change in data arrangement has no effect on the memory access pattern in the activation function. Note that this function is only used in the first feed-forward component of the transformer encoder layer. Due to its element-wise characteristics, it is integrated directly into the feed-forward layer immediately prior to saving the computed values back into the memory. As a result, this particular activation function does not require separate data retrieval from memory.

\section{Experimental Setup and Results}
\label{sec:experiments}

\subsection{Experimental Setup}

We consider BERT-base model~\cite{devlin2018bert}, as a case study. The input matrix for this model has dimensions of 512x768, where 512 represents the sequence length. The Query, Key, and Value matrices for each of the model's 12 heads have dimensions of 768x64, while the feed-forward matrices have a size of 3072.

We used the gem5-X system simulation framework~\cite{qureshi2019gem5} to explore various performance indicators, including execution time and accesses / failures at different levels of the memory hierarchy. gem5-X is an open-source environment extending gem5~\cite{lowe2020gem5} and adding advanced features such as shared guest / host spaces and fine-grained checkpointing. The behavioral models of the accelerators, including the functionality of each defined instruction, are modeled in C++.

Across experiments, we targeted a hardware system that included a single-, dual-, or quad-core processor. The system features 32 KB of L1 cache for instructions and 32 KB of L1 cache for data for each core, 1 MB of L2 cache shared among the cores, and 4 GB of off-chip memory. Its CPU operates at a frequency of 2.3 GHz. Transformer applications run in full-system mode on the Ubuntu 16.04 operating system.

We considered two classes of accelerators: systolic arrays with a kernel size of 16 and 8 elements and SIMD functional units with a kernel size of 16 elements.
In more detail,  we employ the gem5-X systolic matrix model fromin~\cite{amirshahi2023tic}, integrated as a custom functional unit as an example of a  systolic array. In contrast, we focus on the ARM NEON model, included as part of the gem5 framework, as a SIMD accelerator.

\subsection{Impact of BWMA on inference run-time}
\label{sec:results:run-time}
\begin{figure}[t]
  \centering
  \begin{subfigure}{0.48\textwidth}
    \centering
    \includegraphics[width=\linewidth]{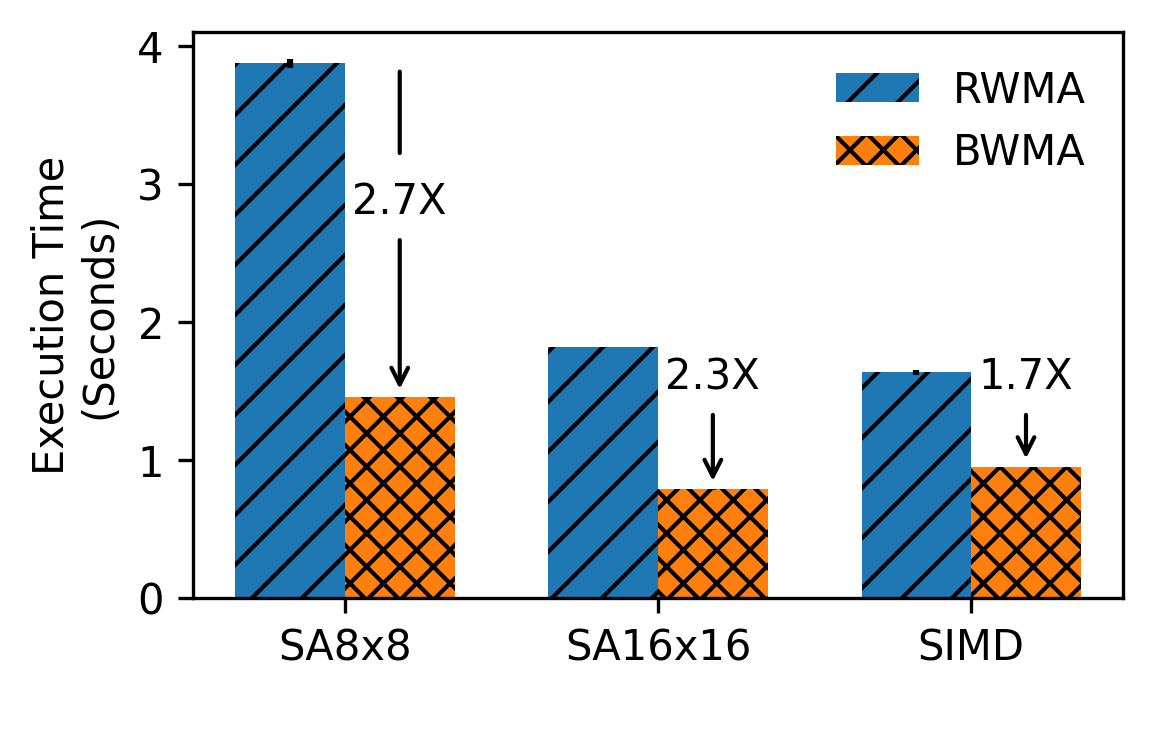}
    \caption{}
    \label{fig:bw_vs_rw_single}
  \end{subfigure}
  \begin{subfigure}{0.48\textwidth}
    \centering
    \includegraphics[width=\linewidth]{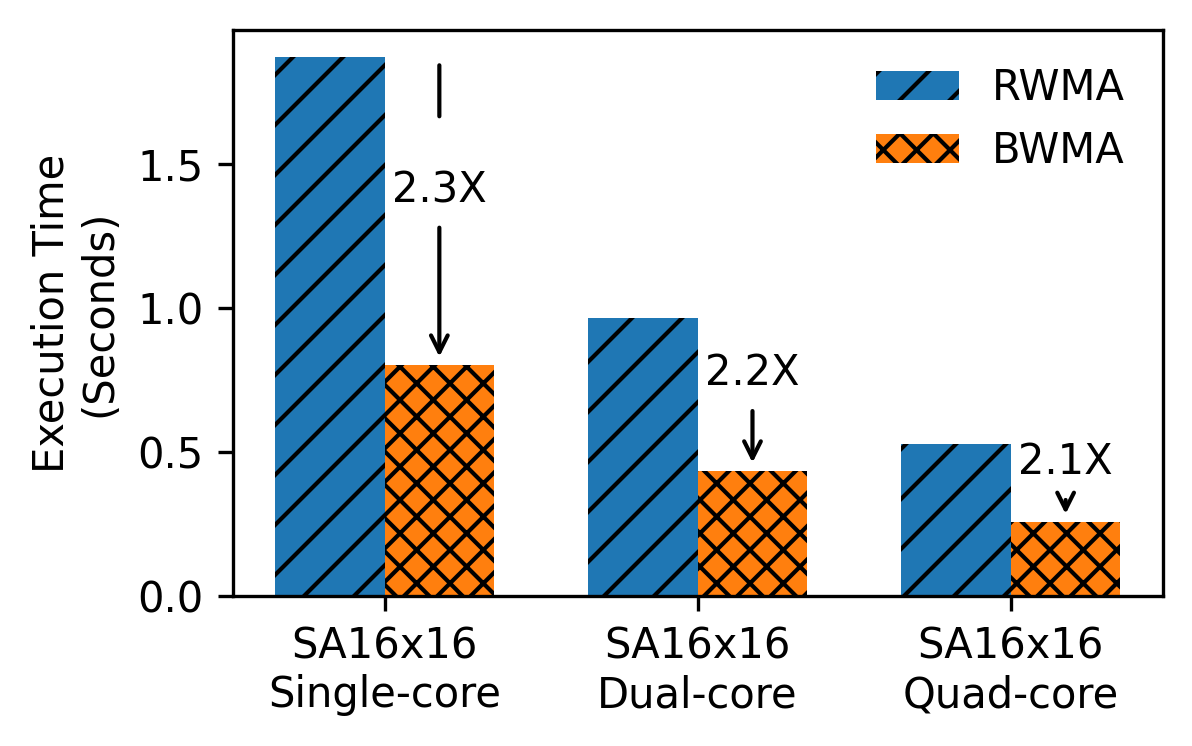}
    \caption{}
    \label{fig:bw_vs_rw_multi}
  \end{subfigure}
  \caption{Execution Time Comparison on BERT model: BWMA demonstrates reduced execution time for transformer models across various (a) hardware accelerators and (b) number of cores.}
  \label{fig:bw_vs_rw}
\end{figure}

Figure~\ref{fig:bw_vs_rw} presents the execution time of a BERT transformer layer when employing various hardware accelerators, such as Systolic Array with a block size of 8x8~(SA8x8), SA16x16, and SIMD on a single-core CPU. As depicted in the figure,  BWMA considerably reduces the execution time of the transformer model in comparison to RWMA, by up to 2.7X in the SA8x8 case. 

In cases where the accelerator has multi-core processing units, BWMA is still very effective in accelerating  the transformer model execution. Figure~\ref{fig:bw_vs_rw_multi} shows the inference run-time for a single-, dual-, quad-core system, where each core embeds dedicated 16x16 SAs. As shown in the results, BWMA can decrease the run-time compared to RWMA irrespectively of the number of cores. Furthermore, this figure shows that the inference run-time for a single-core BWMA is even less than that of a dual-core RWMA, highlighting that optimizing the memory arrangement (which has no hardware cost) can be more effective than duplicating the system resources.  

\begin{figure}[t]
  \centering
  \begin{subfigure}{0.55\textwidth}
    \centering
    \includegraphics[width=\linewidth]{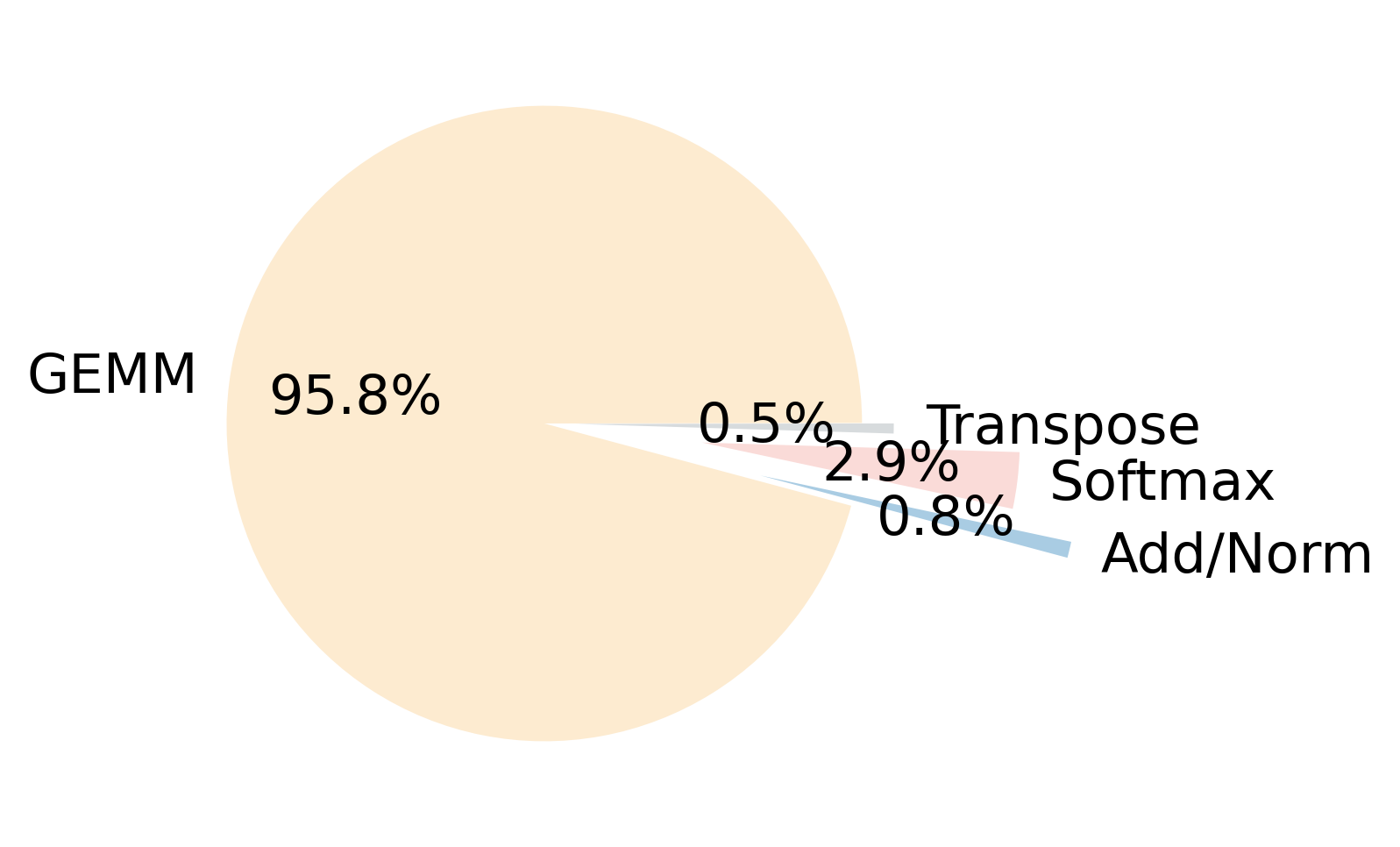}
    \caption{RWMA}
    \label{fig:rw-pie}
  \end{subfigure}
  \begin{subfigure}{0.44\textwidth}
    \centering
    \includegraphics[width=\linewidth]{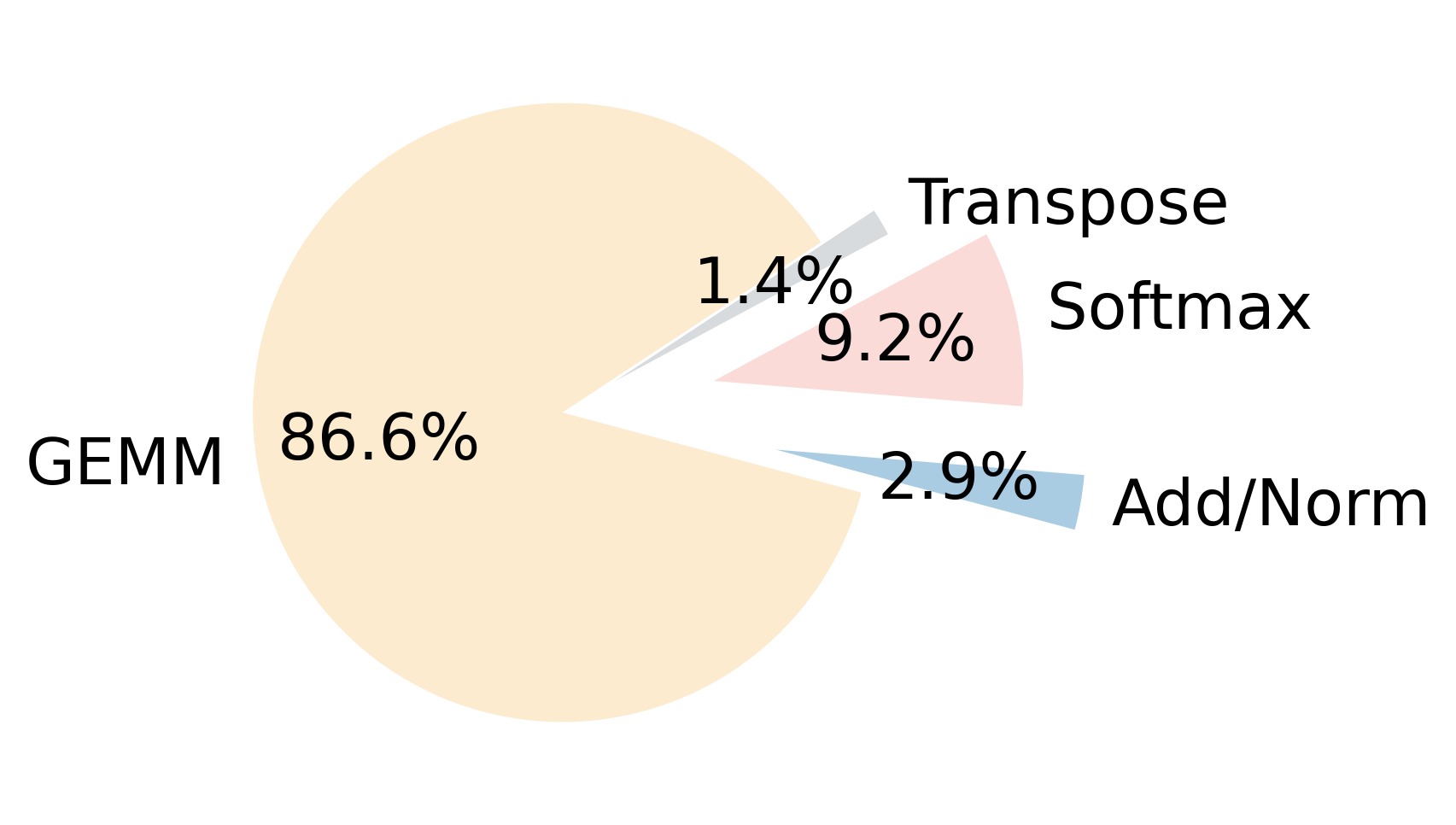}
    \caption{BWMA}
    \label{fig:bw-pie}
  \end{subfigure}
  \caption{Execution Time Distribution in SA16x16 in RWMA and BWMA on a single-core system. The area of the pie charts is proportional to the measured inference times (2.3x smaller for BWMA). The run-time of GEMM operation is the majority, even in the presence of hardware acceleration.}
  \label{fig:pie}
\end{figure}

A more detailed view of the run-time of RWMA inference for a SA16x16 case with a single-core system is provided in Figure~\ref{fig:rw-pie}. The figure depicts the proportion of execution time for each individual component. It can be noticed that the non-GEMM components (Transpose, Softmax, and Add/Norm) add up to only 4.2\% of the execution time of a whole transformer layer. As discussed in Section~\ref{sec:bw_mem:non-gemm}, GEMM operations within these components benefit from memory arrangement, while the non-GEMM components are subject to overhead.

Figure~\ref{fig:bw-pie} similarly displays the proportion of execution time in a similar way for the BWMA case. In BWMA, the proportion of GEMMs execution time decreases, while the non-GEMM components' run-time increases. Despite this last effect, non-GEMMs only account for 13.5\% of the execution time, with the GEMM portion still comprising the majority. Hence, as the decrease in GEMM execution time largely overcompensates for the increased overhead in the non-GEMM components, ultimately resulting in large performance improvements at the application level.

\subsection{Impact on memory accesses}

\begin{figure}[t]
    \centering
    \includegraphics[width=0.74\linewidth]{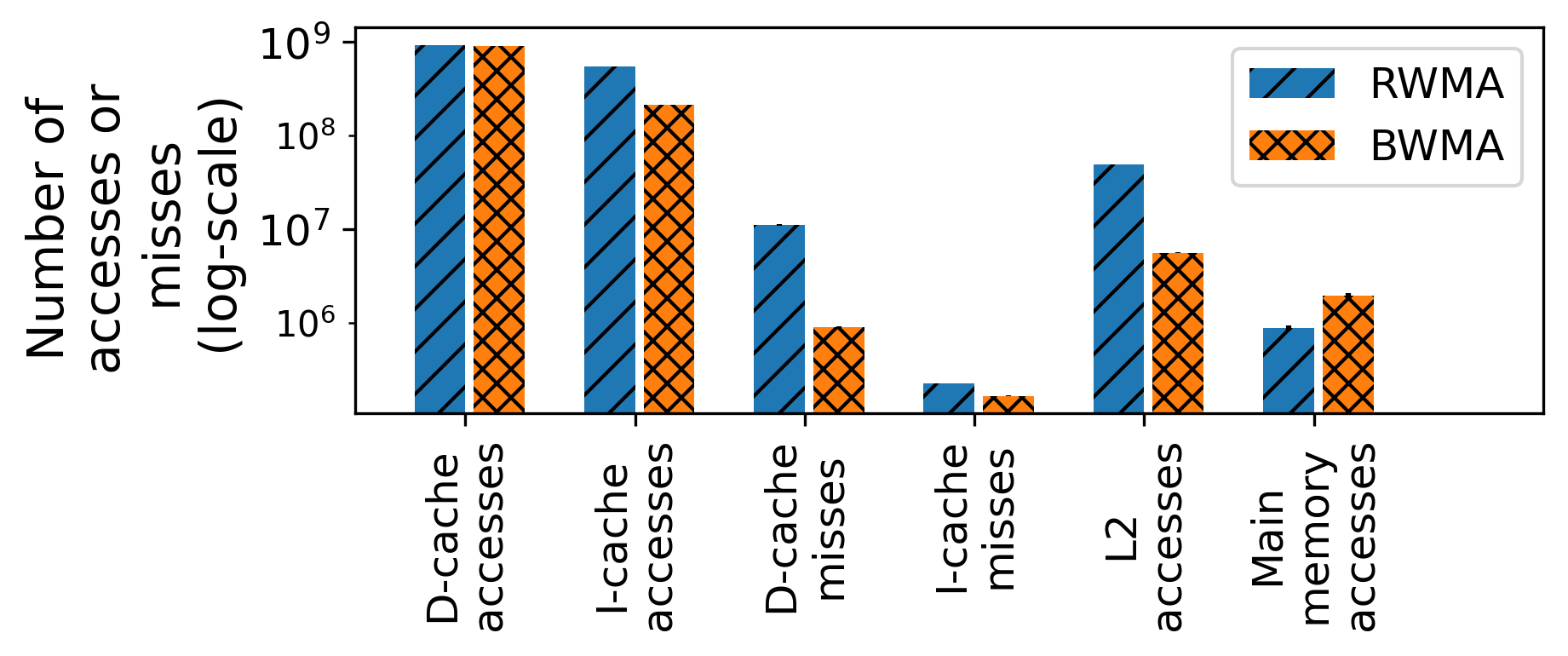}
    \caption{Memory Access Comparison: BWMA results in significantly reduced L2 cache accesses and D-cache misses compared to RWMA, showcasing improved data reutilization and overall performance.}
    \label{fig:bw_mem}
\end{figure}

The speed-up outlined above is caused by the better use of the memory hierarchy of BWMA with respect to RWMA when performing tiled GEMMs. To further illustrate this aspect, Figure~\ref{fig:bw_mem} displays  memory accesses and memory misses for both RWMA and BWMA for a single-core SA16x16. The data are on a logarithmic scale and encompass cache levels as well as main memory. 

For both methods, the number of data accesses requested by the processor is almost the same. Thus, the number of L1 data cache (D-cache) accesses remains nearly constant. In contrast, L1 instruction cache (I-cache) accesses are higher in the case of RWMA, because the data in each tile have to be explicitly indexed. This is not the case for BWMA, where the values belonging to each tile are stored in contiguous memory locations. However, although the I-cache accesses for the RWMA are higher, these are well served by the L1 I-cache, with comparatively few misses.

In contrast,  the number of L2 cache accesses is significantly different in RWMA compared to BWMA. Indeed,  BWMA leads to a higher data reuse in L1 with 12.3 times fewer L1 D-cache misses and, therefore, much fewer L2 accesses. Given the difference in access latency between L1 (2 cycles) and L2 (20 cycles) caches, the BWMA approach results in much better overall performance.

\section{Conclusion}
\label{sec:conclusion}
This paper has demonstrated the effectiveness of aligning memory data arrangements with hardware accelerators to enhance the performance of transformer models. Our proposed arrangement technique, called BWMA, significantly optimizes memory access patterns, which leads to substantial improvements in processing speed and efficiency. In particular, implementing BWMA in transformer models has shown up to a 2.8x increase in speed for single-core accelerator systems. Also, we have shown that while there is a slight increase in runtime for non-GEMM components due to BWMA, this is negligible compared to the overall computational benefits. The BWMA method is scalable and applicable to various hardware architectures and can be effectively utilized in multi-core systems, hence providing a versatile solution for memory data arrangement in diverse computational environments.


\bibliography{oasics-v2021-sample-article}

\begin{thebibliography}{10}

\bibitem{amirshahi2023tic}
Alireza Amirshahi, Joshua Alexander~Harrison Klein, Giovanni Ansaloni, and David Atienza.
\newblock Tic-sat: Tightly-coupled systolic accelerator for transformers.
\newblock In {\em Proceedings of the 28th Asia and South Pacific Design Automation Conference}, pages 657--663, 2023.

\bibitem{devlin2018bert}
Jacob Devlin, Ming-Wei Chang, Kenton Lee, and Kristina Toutanova.
\newblock Bert: Pre-training of deep bidirectional transformers for language understanding.
\newblock {\em arXiv preprint arXiv:1810.04805}, 2018.

\bibitem{dosovitskiy2020image}
Alexey Dosovitskiy, Lucas Beyer, Alexander Kolesnikov, Dirk Weissenborn, Xiaohua Zhai, Thomas Unterthiner, Mostafa Dehghani, Matthias Minderer, Georg Heigold, Sylvain Gelly, et~al.
\newblock An image is worth 16x16 words: Transformers for image recognition at scale.
\newblock {\em arXiv preprint arXiv:2010.11929}, 2020.

\bibitem{eggermann2023sixteen}
Gr{\'e}goire~Axel Eggermann, Marco~Antonio Rios, Giovanni Ansaloni, David Atienza~Alonso, and Sani Nassif.
\newblock A 16-bit floating-point near-sram architecture for low-power sparse matrix-vector multiplication.
\newblock In {\em VLSI SoC}, 2023.

\bibitem{ferry2022increasing}
Corentin Ferry, Tomofumi Yuki, Steven Derrien, and Sanjay Rajopadhye.
\newblock Increasing fpga accelerators memory bandwidth with a burst-friendly memory layout.
\newblock {\em IEEE Transactions on Computer-Aided Design of Integrated Circuits and Systems}, 2022.

\bibitem{herrero2006using}
Jos{\'e}~R Herrero and Juan~J Navarro.
\newblock Using non-canonical array layouts in dense matrix operations.
\newblock In {\em International Workshop on Applied Parallel Computing}, pages 580--588. Springer, 2006.

\bibitem{hsu2021hubert}
Wei-Ning Hsu, Benjamin Bolte, Yao-Hung~Hubert Tsai, Kushal Lakhotia, Ruslan Salakhutdinov, and Abdelrahman Mohamed.
\newblock Hubert: Self-supervised speech representation learning by masked prediction of hidden units.
\newblock {\em IEEE/ACM Transactions on Audio, Speech, and Language Processing}, 29:3451--3460, 2021.

\bibitem{khoram2017challenges}
Soroosh Khoram, Yue Zha, Jialiang Zhang, and Jing Li.
\newblock Challenges and opportunities: From near-memory computing to in-memory computing.
\newblock In {\em Proceedings of the 2017 ACM on International Symposium on Physical Design}, pages 43--46, 2017.

\bibitem{kim2023full}
Sehoon Kim, Coleman Hooper, Thanakul Wattanawong, Minwoo Kang, Ruohan Yan, Hasan Genc, Grace Dinh, Qijing Huang, Kurt Keutzer, Michael~W Mahoney, et~al.
\newblock Full stack optimization of transformer inference: a survey.
\newblock {\em arXiv preprint arXiv:2302.14017}, 2023.

\bibitem{lam1991cache}
Monica~D Lam, Edward~E Rothberg, and Michael~E Wolf.
\newblock The cache performance and optimizations of blocked algorithms.
\newblock {\em ACM SIGOPS Operating Systems Review}, 25(Special Issue):63--74, 1991.

\bibitem{li2020ftrans}
Bingbing Li, Santosh Pandey, Haowen Fang, Yanjun Lyv, Ji~Li, Jieyang Chen, Mimi Xie, Lipeng Wan, Hang Liu, and Caiwen Ding.
\newblock Ftrans: energy-efficient acceleration of transformers using fpga.
\newblock In {\em Proceedings of the ACM/IEEE International Symposium on Low Power Electronics and Design}, pages 175--180, 2020.

\bibitem{lowe2020gem5}
Jason Lowe-Power, Abdul~Mutaal Ahmad, Ayaz Akram, Mohammad Alian, Rico Amslinger, Matteo Andreozzi, Adri{\`a} Armejach, Nils Asmussen, Brad Beckmann, Srikant Bharadwaj, et~al.
\newblock The gem5 simulator: Version 20.0+.
\newblock {\em arXiv preprint arXiv:2007.03152}, 2020.

\bibitem{peng2021accelerating}
Hongwu Peng, Shaoyi Huang, Tong Geng, Ang Li, Weiwen Jiang, Hang Liu, Shusen Wang, and Caiwen Ding.
\newblock Accelerating transformer-based deep learning models on fpgas using column balanced block pruning.
\newblock In {\em 2021 22nd International Symposium on Quality Electronic Design (ISQED)}, pages 142--148. IEEE, 2021.

\bibitem{qi2021accelerating}
Panjie Qi, Edwin Hsing-Mean Sha, Qingfeng Zhuge, Hongwu Peng, Shaoyi Huang, Zhenglun Kong, Yuhong Song, and Bingbing Li.
\newblock Accelerating framework of transformer by hardware design and model compression co-optimization.
\newblock In {\em 2021 IEEE/ACM International Conference On Computer Aided Design (ICCAD)}, pages 1--9. IEEE, 2021.

\bibitem{qureshi2019gem5}
Yasir~Mahmood Qureshi, William~Andrew Simon, Marina Zapater, David Atienza, and Katzalin Olcoz.
\newblock Gem5-x: A gem5-based system level simulation framework to optimize many-core platforms.
\newblock In {\em 2019 Spring Simulation Conference (SpringSim)}, pages 1--12. IEEE, 2019.

\bibitem{vaswani2017attention}
Ashish Vaswani, Noam Shazeer, Niki Parmar, Jakob Uszkoreit, Llion Jones, Aidan~N Gomez, {\L}ukasz Kaiser, and Illia Polosukhin.
\newblock Attention is all you need.
\newblock {\em Advances in neural information processing systems}, 30, 2017.

\bibitem{yang2022efa}
Xin Yang and Tao Su.
\newblock Efa-trans: An efficient and flexible acceleration architecture for transformers.
\newblock {\em Electronics}, 11(21):3550, 2022.

\bibitem{you2023vitcod}
Haoran You, Zhanyi Sun, Huihong Shi, Zhongzhi Yu, Yang Zhao, Yongan Zhang, Chaojian Li, Baopu Li, and Yingyan Lin.
\newblock Vitcod: Vision transformer acceleration via dedicated algorithm and accelerator co-design.
\newblock In {\em 2023 IEEE International Symposium on High-Performance Computer Architecture (HPCA)}, pages 273--286. IEEE, 2023.

\bibitem{zhong2023transformer}
Juan Zhong, Zheng Liu, and Xi~Chen.
\newblock Transformer-based models and hardware acceleration analysis in autonomous driving: A survey.
\newblock {\em arXiv preprint arXiv:2304.10891}, 2023.

\end{thebibliography}

\end{document}